\documentclass[prl,twocolumn,superscriptaddress,preprintnumbers,amssymb,epsfigs]{revtex4}

\usepackage{graphicx}

\usepackage{dcolumn}

\usepackage{bm}

\usepackage{wasysym}
\usepackage{amsthm,amsfonts,epsfig,verbatim}

\def\beq{\begin{equation}}
\def\eeq{\end{equation}}
\def\beqn{\begin{eqnarray}}
\def\eeqn{\end{eqnarray}}
\def\be{\begin{equation}}
\def\ee{\end{equation}}
\def\ba#1{\begin{array}{#1}}
\def\ea{\end{array}}
\def\bn{\begin{enumerate}}
\def\en{\end{enumerate}}

\def\l{\left}

\def\H{\mathcal{H}}
\def\summ{\sum\limits}

\def\l{\left}
\def\rr{\right}

\newcommand{\tr}{{\rm tr}}

\newcommand{\D}{{\rm d}}

\begin{document}
\title{Towards measuring Entanglement Entropies in Many Body Systems.}

\author{Israel Klich  }%
\author{Gil Refael }%
\affiliation{ Department of Physics, California Institute of
  Technology, MC 114-36 Pasadena, CA 91125}
\author{Alessandro Silva }%
\affiliation{ Abdus Salam ICTP, Strada Costiera 11, 34100 Trieste,
Italy}

\begin{abstract}
We explore the relation between entanglement entropy of quantum many body systems and
the distribution of corresponding, properly selected, observables. Such a relation is
necessary to actually {\it measure} the entanglement entropy. We show that in general,
the Shannon entropy of the probability distribution of certain symmetry observables
gives a lower bound to the entropy. In some cases this bound is saturated and directly
gives the entropy. We also show other cases in which the probability distribution
contains enough information to extract the entropy: we show how this is done in several
examples including BEC wave functions, the Dicke model, XY spin chain and chains with
strong randomness.
\end{abstract}
\maketitle

Entanglement entropy was first considered as a source of quantum corrections to the
entropy of a black hole \cite{Bombelli}. Since then its study expanded considerably in
extent and purpose. Today its evaluation allows an estimate of the effectiveness of
physical systems as quantum computers, since the entanglement entropy characterizes the
resources in our disposal to perform computations. As importantly, the entanglement
entropy is emerging as a tool in the field of many body systems. Close to quantum phase
transitions and at critical points it was shown to exhibit universal properties
\cite{Cardy,Goldbart}. In particular, in one dimensional conformally invariant systems
the entanglement entropy is a measure of the central charge of the underlying conformal
field theory \cite{Vidal}.

Consider a system partitioned into two parts $A$ and $B$. The entanglement entropy
${\cal S}_E$ in the quantum state $\rho=\l|\Psi\rangle\langle\Psi\rr|$ evaluates how
many qubits in  $A$ are determined by (or entangled with) qubits in $B$. It is defined
as the von Neumann entropy ${\cal S}_{E}=-{\rm Tr}[\rho_A\;\log_{2}(\rho_A)]$ of the
reduced density matrix $\rho_A={\rm Tr_B[\rho]}$.

As it stands, the notion of entanglement entropy is  rather
abstract. Indeed, to the best of our knowledge, despite the large
body of literature dedicated to the study of entanglement
entropies in many body systems, the actual physical possibility of
\it measuring \rm this quantity, and so investigating
experimentally its properties (such as scaling behavior), has not
been addressed. This is the main motivation of this Letter.

It is clear that, given the entire distribution of correlation functions and
observables one may always recover the density matrix, and hence ${\cal S}_E$. A step
in this direction was taken in Ref. \cite{Wu Sarandy Lidar Sham}, where it was
emphasized that a quantum state may be reconstructed from the set of susceptibilities
with respect to a large enough class of external potentials via the Hohenberg-Kohn
theorem. In Ref. \cite{Wu Sarandy Lidar Sham}, this was used to calculate the linear
entanglement entropy and negativity of at most 4 dimensional subspace (for example two
spins) with a large system. It is, however, unclear how this can be implemented to find
the entanglement between two large subsystems where the entire density matrix cannot be
realistically measured.

In this letter we show that, given a quantum many body system, one may always identify
a class of natural observables, whose fluctuations can be related to the entanglement
entropy. After some general considerations, we provide several examples of systems
where the entanglement entropy can be estimated, or directly extracted, from the
probability distribution $P(x)$ of the possible outcomes $x$ of such observables. This
connection will take different forms in different problems.

A key ingredient of our analysis consists in considering directly, instead of ${\cal
S}_E$, the so called "measurement entropy" $S[\hat{O}]$ of properly chosen observables
$\hat{O}$, i.e. the Shannon entropy $S[\hat{O}]=-\sum_{x} P(x) \log P(x)$ associated
with the probability distribution $P(x)$ of the outcomes $x$ of $\hat{O}$
\cite{Balian}. In classical systems this quantity, measurable by definition, is always
lower then the overall entropy. On the other hand, in quantum systems it can in general
be either larger or smaller than the entanglement entropy \cite{note1}. Below we select
a class of local observables such that $S[\hat{O}]$ provides a lower bound on the
entropy $S_E$. While in some cases the bound is saturated and $S[\hat{O}]=S_E$, in
other cases, where estimates based on $S[\hat{O}]$ do not capture the scaling behavior,
we are nevertheless able to extract ${\cal S}_E$ from the distribution $P(x)$.

The class of observables whose measurement enables us to estimate
the entanglement entropy is constructed as follows. Given a state
$ \psi $ of interest, let us denote by $\cal L$ the set of
observables $\hat{O}=\hat{O}_A\otimes I+I\otimes \hat{O}_B$,
acting locally on $A$ and $B$, for which $\psi$ is an eigenstate.
Note that $\cal L$ is nonempty \cite{note2}. The Schmidt
decomposition of $\psi$ can be written as $\psi=\sum
c^{\alpha}_{i}|\alpha,i>\otimes|s-\alpha,i>$, where $s$ is the
eigenvalue of $\hat{O}$ acting on $\psi$, and such that
$\hat{O}_A|\alpha,i>=\alpha|\alpha,i>$ and
$\hat{O}_B|s-\alpha,i>=(s-\alpha)|s-\alpha,i>$ (here $i$ ranges
over the degeneracy of eigenstates of $\hat{O}_A$ with value
$\alpha$)

One may write the reduced density matrix as $\rho_A=\tr_B
\rho=\sum P_{\alpha}\rho_{\alpha}$, where we have defined
$\rho_{\alpha}={1\over P_{\alpha}}\sum
|c^{\alpha}_{i}|^2|\alpha,i><\alpha,i| $ with $P_{\alpha}=\sum_i
|c^{\alpha}_{i}|^2$. Therefore, for the entanglement entropy one
has
\begin{eqnarray}
{\cal S}_{E} = S[\hat{O}_A]-\sum
P_{\alpha}\tr\rho_{\alpha}\log{\rho_{\alpha}}\geq S[\hat{O}_A],
\label{measure}
\end{eqnarray}
where $S[\hat{O}_A]$ is the measurement entropy associated to the probability
distribution $P_{\alpha}$.

The inequality in Eq. (\ref{measure}) is
  completely general. Interestingly, the equality ${\cal S}_{E}=S[\hat{O}_A]$ is realized if and only if
$\tr\rho_{\alpha}\log{\rho_{\alpha}}=0$ for all $\alpha$, as for example in the case
of no degeneracy of the eigenvalue $\alpha$, or in more interesting cases
where the $\rho_\alpha$'s describe pure states. Such systems do
  indeed exist: we will consider, for example, a BEC like state, a two mode squeezed state, and
the Dicke model. As remarked above the measurement entropy has the advantage of being
directly measurable. By repeated measurements of $\hat{O}_A$ we can recover the
distribution of outcomes, and so extract $S[\hat{O}_A]$. Moreover, we note that ${\cal
S}_{E}=Max_{O\in \cal L}S[\hat{O}_A]$. To see this, choose the operator $\hat{O}_A$
diagonal with non-degenerate eigenvalues $\lambda_i$ in the basis $|i_A\rangle$
appearing in the Schmidt decomposition of $\psi$, and $\hat{O}_B$ diagonal with
eigenvalues $-\lambda_i$ in the basis $|i_B\rangle$, $\psi$ is an eigenstate of
eigenvalue $0$ of $\hat{O}_A\otimes I+I\otimes \hat{O}_B$, and
  $S[\hat{O}_A]={\cal S}_{E}$. Note that in the same way, if we define $S_L[\hat{O}]=\sum
P(x)(1-P(x))$ to be the {\it linear measurement entropy} of $O$,
it obeys $S_L\geq S_L[\hat{O}]$, where $S_L=1-\tr\rho^2$ is the
linear entropy of the system. In the case $S[\hat{O}]={\cal
S}_{E}$ one also immediately has $S_L[\hat{O}]=S_L$.


Intuitively, a natural subset of $\cal L$ is the set of "conserved" operators, i.e.,
sums of local operators which commute with the Hamiltonian of the system and thus are
in $\cal L$. For instance, the total spin operator for spin chains with rotational
symmetry. These are the cases we consider. We remark, however, that since not for every
Hamiltonian the set $\cal L$ necessarily contains such conserved observables
\cite{note3}, in the general case, the appropriate choice of $\hat{O}$ may requires a
more elaborate analysis.

Let us now illustrate the general considerations above with two examples in which the
equality $S[\hat{O}]={\cal S}_E$ is actually satisfied. As a first illustrative
example, consider a BEC condensate of bosons who share a particular Gross-Pitaevskii
wave function $f(x)$, constrained to occupy a box of volume $V$. One can imagine
dividing the box into two parts $A$ and $B$, of volumes $V_A$ and $V_B$ respectively.
The condensate wave function can then be written as
\begin{eqnarray}
\psi&=& {1\over \sqrt{N!}}(u a^{\dag}+\sqrt{1-u^2} b^{\dag})^N|0>
\nonumber \\
&=& \sum_{k=0}^N \sqrt{C^N_k} u^k\;(1-u^2)^{N-k}\mid k \rangle_A
\otimes \mid N-k \rangle_B,
\end{eqnarray}
where $N$ is the number of bosons, $C^N_k$ is the binomial coefficient, and
$a^{\dag},b^{\dag}$ create a particle in $A,B$ respectively, i.e., $a^{\dag}={1\over
u}\int_{x\in A}f(x)\psi^{\dag}(x)\D x$ and $b^{\dag}={1\over \sqrt{1-u^2}}\int_{x\in
B}f(x)\psi^{\dag}(x)\D x$, with $u^2=\int_{x\in A} |f(x)|^2\D x$. Note that $\psi$ is
an eigenstate of the particle number operator $\hat{N}=\int_{x\in A}\D
x\psi^{\dag}_x\psi_x+\int_{x\in B}\D x \psi^{\dag}_x\psi_x$. We may consider just the
subspace consisting of applications of $a^{\dag}$ and $b^{\dag}$ to the vacuum, and
ignore other modes. In this subspace there is no degeneracy since every occupation
number appears only once and so we have simply ${\cal
S}_{E}=S[a^{\dag}a]=-\sum\;P_k\;\log_2(P_k)$, with $P_k=C^N_k (u^2)^{N-k}(1-u^2)^{k}$.
Consequently the measurement of entropy of particle number in $A$ in this case is
exactly the entanglement entropy. Assuming for simplicity $V_B>>V_A$, and taking $f(x)$
as uniform [$f(x)=1/ \sqrt{V_A+V_B}$] then
in the limit $N \gg 1$, one obtains ${\cal S}_{E} \simeq 1/2\;\log[N u^2(1-u^2)]$,
which, in the thermodynamic limit ($N \rightarrow +\infty$, keeping $\nu=N/V$
constant), simplifies to ${\cal S}_{E} \simeq 1/2\;\log( V_A \nu)$.


A similar example of entangled state, but this time of two
distinct degrees of freedom, is a two-mode squeezed state
\begin{eqnarray}\label{twomode}
\psi &=& \exp\left[\xi a_1^{\dag} a^{\dag}_2- \bar{\xi} a_1
a_2 \right] | 0 > \nonumber \\
&=& {1\over\cosh[\eta]}\sum_{n=0}^{+\infty} e^{i n \Phi}
\left[\tanh(\eta) \right]^n \mid n \rangle_1 \otimes \mid n
\rangle_2,
\end{eqnarray}
where $\xi=\eta \;e^{i\Phi}$ is the squeezing parameter, and $a_1$, $a_2$ are the
annihilation operators relative to the two modes. Identifying $A$ and $B$ with the two
modes, $\psi$ is an eigenstate of the sum of local operators $\hat{O}= \hat{n}_1
\otimes I + I \otimes (- \hat{n}_2)$. Since the eigenstates of $n_1$ and $n_2$ are non
degenerate the entanglement entropy is ${\cal S}_{E}=S[\hat{n}_1]$. In particular,
${\cal S}_{E}=\log_2[1+\bar{n}]+\log_2\left[(1+\bar{n})/ \bar{n} \right]\bar{n}$.
where $\bar{n}=[\sinh(\eta)]^2$ is the average number of photons
per mode. In the limit $\eta \rightarrow +\infty$, one obtains
$S_{12} \approx \log_2[\bar{n}]$.

Squeezed states can be generated in a number of physical situations. A particularly
interesting realization is obtained when a collection of two level atoms (subsystem
$A$) interacts with a single mode of the EM field (subsystem $B$): the Dicke model. In
turn, the Dicke model can be realized in a number of ways, both using traditional
cavity QED, as well as employing solid state circuits [circuit QED]\cite{Girvin}.
Assuming for simplicity the photon mode to be in resonance with the atoms, the
Hamiltonian is
\begin{eqnarray}
H=\omega_0 \hat{J}_z +\omega_0 a^{\dag} a + {\lambda \over \sqrt{2
j}} (a^{\dag}+a)\;(\hat{J}_{+}+\hat{J}_{-}),
\end{eqnarray}
where $\hat{J}_k=\sum_{i=1}^{2j} s^i_{k}\;\;[k=\pm,z]$ are
collective operators describing the dynamics of the collection of
$2j$ two level atoms. Working in the subspace where, at
$\lambda=0$, all atoms are in the ground state,
in the large $j$ limit, one can conveniently use the
Holstein-Primakoff representation $\hat{J}_z=b^{\dag}b-j$,
$\hat{J}_{+}=b^{\dag}\;\left(\sqrt{2j-b^{\dag}b}\right)$,
$\hat{J}_{-}=\left(\sqrt{2j-b^{\dag}b}\right)\;b$, where $b$ are
bosonic modes~\cite{Brandes}. In the $j \rightarrow +\infty$
limit, one obtains
$H = \omega_0 b^{\dag}b +\omega_0 a^{\dag} a + \lambda
(a^{\dag}+a)\;(b^{\dag}+b)-j\omega_0$. This Hamiltonian describes
two coupled harmonic oscillators. In particular, as a function of
coupling $\lambda$, the Hamiltonian has a Quantum Critical Point
at $\lambda_c=\omega_0/2$, where the ground state symmetry with
respect to parity is spontaneously broken. Setting $\lambda \leq
\lambda_c$, it is convenient to introduce the operators $x={1\over
\sqrt{2\omega_0}}(a+a^{\dag})$, $y={1\over
\sqrt{2\omega_0}}(b+b^{\dag})$, $p_x=i\sqrt{{\omega_0\over 2}}
(a^{\dag}-a)$ and $p_y=i\sqrt{{\omega_0\over 2}} (b^{\dag}-b)$.
Writing down the Hamiltonian in terms of these operators, and
making the rotation $q_1=(x+y)/\sqrt{2}$, $q_2=(x-y)/\sqrt{2}$ one
obtains a quadratic Hamiltonian, the spectrum being that of two
independent harmonic oscillators of frequencies $K\;e^{\pm
4\eta}$, where $K=\sqrt{\omega_0^4-4\lambda^2 \omega_0^2}$, and
$e^{4\eta}=\sqrt{(\omega_0^2-2\lambda
\omega_0)/(\omega_0^2+2\lambda \omega_0)}$. The ground state can
be written as
\begin{eqnarray}
\psi =\left[{K \over \pi^2} \right]^{1/4} e^{ -\frac{1}{2} \left[
e^{-2\eta} (K^{1/4} q_1)^2 + e^{2\eta} (K^{1/4} q_2)^2 \right]},
\end{eqnarray}
Introducing $\Phi_n(x)$, the eigenstates of the operator $H_x(K)=(p_x^2+K\;x^2)/2$, it
is holds that
\begin{eqnarray}
\psi ={1\over \cosh(\eta)} \sum_{n=0}^{+\infty} \left[ \tanh(\eta)
\right]^n \Phi_n(x) \Phi_n(y).
\end{eqnarray}
Therefore, one immediately finds that, defining the operator $\hat{O}=H_x(K)\otimes I +
I \otimes (-H_y(K))$, the ground state of the model is an eigenstate with zero
eigenvalue. In addition, ${\cal S}_{E}=S[H_x(K)]$. It is interesting to notice that,
for $\lambda \ll \lambda_c$, one can approximate $H_{x} \simeq
\omega_0\;(a^{\dag}a+{1\over 2})$, while for $\lambda \simeq \lambda_c$ one has $H_{x}
\simeq p_x^2/2$. In analogy with the previously considered two mode squeezed state, for
$\lambda \simeq \lambda_c$ the entanglement entropy is given by ${\cal S}_{E} \simeq
2\;\log_2(\sinh(\eta)) \approx 1/4\;\log[{\omega_0\over
|\lambda-\lambda_c|}]$~\cite{Brandes}.

The lower bound on ${\cal S}_E$ provided by the measurement
entropy has to be considered with care, in particular in
extracting the scaling of ${\cal S}_E$ with subsystem size $L_A$
in one [or higher] dimensional systems. In these physical
situations, it is convenient to choose the operators $\hat{O}$ as
corresponding to {\it extensive} observables, such as the total
magnetization, or particle number. In this case, one expects the
distribution of measurements outcomes $x$ of $\hat{O}$ to be
Gaussian in the limit of large subsystem, i.e. $P(x)\rightarrow
{1\over\sqrt{2\pi\sigma^2}}e^{-(x-L)^2/{2\sigma^2}}$ for
$L_A\rightarrow\infty$, For such observables we have
$S[\hat{O}]=1+{1\over 2}\log (2\pi\sqrt{<\Delta x^2>})$.
More generally, given the variance $\sqrt{<\Delta x^2>}$, the
formula above gives always the maximal measurement entropy of a
continuous variable, as one can easily checked by a variational
argument. In other words, the variance provides an upper bound to
the measurement entropy.


While for such extensive operators one immediately obtains $S[\hat{O}] \approx  \log
(\sqrt{<\Delta x^2>}) \leq {\cal S}_E $, in many cases this inequality is of limited
use, since, while $S[\hat{O}]$ scales at most logarithmically with the variance, the
entanglement entropy scales in fact as the variance itself and not its logarithm. This
implies that, in many cases, the variance is a useful entanglement measure. Indeed the
variance is decreased in any local physical POVM measurements respecting
super-selection rules \cite{SchuchVerstraeteCirac}.



To demonstrate this, let us now  consider the $XY$ spin chain \cite{Vidal} and cases of
non-interacting fermions \cite{Peschel}. Under the Jordan-Wigner mapping one can treat
the $XY$ spin chain as a special case of non-interacting fermions on a $1D$ tight
binding model, with no external potentials. The role of the total $S^z_A$ observable
$\sum_{i\in A} S^z_{i}$ is played by the fermion number operator $\sum_{i\in
A}c^{\dag}_i c_i$. The entanglement of free fermions is particularly interesting as it
exhibits scaling violating area law in higher dimensions \cite{Gioev Klich,Wolf}.
Indeed, a formula for the entropy scaling was presented in \cite{Gioev Klich} in $d$
dimensions, and was recently checked numerically for certain models in 2$d$ and 3$d$
\cite{num_check}. The method presented here is independent of dimension and is also
valid for fermions in the presence of external potentials. For a non-interacting
fermion system the ground state is given by filling low energy modes $\phi_i$. The
entropy of a region $A$ may be expressed as \cite{klich}
\begin{eqnarray}\label{entropy}
  {\cal S}_{E}({L})=-\tr [M\log M+(1-M)\log(1-M)]
\end{eqnarray}
where $M$ is a matrix with elements $M_{ij}=<\phi_i|P_A|\phi_j>$, where $P_A$ is
projection on the region $A$.

The distribution of fermion number may be extracted from the moment generating
function $\chi(\lambda)=\sum P_{\alpha}e^{i\lambda \alpha}=\det(1-M+Me^{i\lambda})$
\cite{klich}. Such functions appear in the theory of Quantum optics, where one studies
the distribution of detected photons. More recently the analog of $\chi$ was introduced
in the theory of quantum transport in mesoscopic systems and is referred to as the full
counting statistics of fermions \cite{Levitov}.

The cummulants of fermion number (or, equivalently $S_z$) in region $A$ are given by
derivatives of $\log\chi$ at $\lambda=0$. Taking derivatives give expressions of the
form $\ll(\delta S_z)^n\gg=(-i)^n\partial_{\lambda}^n\chi|_{\lambda=0}= Q_{n,l} \tr
M^l$.

Now $\tr M^l$ can be extracted from the cummulants by inverting $Q$, i.e. $\tr
M^l=Q^{-1}_{l,n}\ll(\delta S_z)^n\gg$. The matrix $Q^{-1}$ is lower triangular (i.e.
$Q^{-1}_{l,n}=0$ for $l>n$) and may be calculated to any desired order. For example,
$\tr M=\ll\delta S_z\gg$, $\tr M^2=\ll\delta S_z\gg-\ll\delta S_z^2\gg$, $\tr
M^3=\ll\delta S_z\gg-{3\over2}\ll\delta S_z^2\gg+{1\over 2}\ll\delta S_z^3\gg$.

We may write the entropy as the series, convergent whenever the eigenvalues of $M$ are
away from $0$ and $1$
\begin{eqnarray}&
{\cal S}_{E}=-\sum_{n=2}^\infty {1\over n(n-1)}\tr (M^n-1+(1-M)^n)=
\\ \nonumber &
=-\sum_{n=2}^\infty {1\over n(n-1)} (Q^{-1}_{l,n}\ll(\delta
S_Z)^n\gg\\ \nonumber & +\sum_{k=1}^{n}C^n_k
Q^{-1}_{k,j}(-1)^j\ll(\delta S_Z)^j\gg).
\label{expand}
\end{eqnarray}
For any {\it finite} matrix $M$ this series converges as each term is smaller then
$\dim M$, it is easy to check that in fact $ {\cal S}_{E}\leq\log 2\dim M$ as should
be. One must note however that convergence may be slow, depending on the eigenvalues of
$M$ close to unity or zero. Fortunately, all the terms in the series are positive, so
the lower bound improves by adding more terms. In fact, the first non-vanishing
contribution, namely $n=2$ in the series corresponds to the variance, and turns out to
scale in the same way as entropy in the free fermion case \cite{Gioev Klich}.

Remarkably, the coefficients $(Q^{-1})_{l,n}$ appearing are universal - i.e. will fit
any non-interacting fermion theory, independent of its eigenmodes. One possibility is
to measure cold fermionic atoms in a trap; by measuring the histogram of particle
number in a certain region of the trap, one could estimate the entanglement entropy
using Eq. (\ref{expand}) regardless of the particular trap details.


The intimate relation between bipartite entanglement and fluctuations of a conserved
quantity is not restricted to non-interacting systems, but extends also to the strongly
disordered spin-1/2 XXZ model: \be \H=\summ_i J_i\l (\hat{S}^x_i\cdot\hat{S^x}_{i+1}+
\hat{S}^y_i\cdot\hat{S^y}_{i+1}+\lambda \hat{S}^z_i\cdot\hat{S^z}_{i+1}\rr) \label{xxz}
\ee with $-1/2\le\lambda\le 1$ and the $J_i$ are positive and randomly distributed.
This interesting class of interacting random 1-d theories was recently discussed in
Ref. \cite{Refael Moore}, where it was shown that the bipartite entanglement entropy of
a segment of length $L$ with the rest of the chain is ${\cal S}_{E}(L)=\frac{1}{6}\ln
2\log_2 L$. The ground state of the Hamiltonians in Eq. (\ref{xxz}) consists of a
frozen liquid of valence bonds, which is referred to as the random singlet phase
\cite{BhattLee,FisherDoty,DSF95}.

The Hamiltonian (\ref{xxz}) only commutes with $\hat{S}^z_{total}=\summ_i\hat{S}^z_i$
(note, however, that its ground state has a  full rotational symmetry). Therefore
$\hat{S}_A=\summ_{i\in A}\hat{S}^z_{i}$ is the operator of choice for estimating the
entanglement between part A and the rest of the chain. In the random singlet phase
there are two types of singlets: (a) $N_{AB}$ singlets connecting between A and B, (b)
$N_{AA}+N_{BB}$ singlets connecting sites in A to other sites in A, or sites in B to
other sites in B. As explained in Ref. \cite{Refael Moore}, each singlet contributes 1
to the entropy, and therefore: ${\cal S}_E=N_{AB}$. In addition, each singlet
contributes $1/4$ to the variance of $\hat{S}_A$. In this case indeed we have the
relation: \be {\cal S}_{E}=4 \langle \Delta \hat{S}_A^2\rangle=N_{AB}. \label{var} \ee
Note that Eq. (\ref{expand}) still applies in this case where the full counting
statistics of the z-direction spin becomes $\chi(\lambda)=\sum P_{\alpha}e^{i\lambda
  \alpha}=\det((1-M)e^{-i\lambda/2}+Me^{i\lambda/2})$, with $M$ now
being a diagonal matrix with entries $1/2$ for each singlet of type
(a), and $0$ otherwise.

Eq. (\ref{var}) raises the question whether there is an even more direct relationship
between variance of conserved, quantized, quantities and entanglement in the context of
spin chains. We considered this for resonating valance bond (RVB) states of six spins
with varying weights for each singlet configuration, and found that although in most of
the Hilbert space $var S^z_A\le {\cal S}_{E}$, a small region exists where this
inequality is violated. This is associated with the formation of strong ferromagnetic
correlations. It is possible that by putting a few restrictions on a RVB state one can
prove a general relation. Nevertheless, the variance of a conserved quantity of a
quantized object can be thought of as an ad-hoc measure of entanglement.

The interest in entanglement is wide spread due to the prospect of engineering and
controlling entangled states in which two [or more] microscopic objects, although
separated by a macroscopic distance, display quantum correlations. This gives even more
urgency to understanding how to measure entanglement {\it quantitatively}. To determine
whether they have such exotic states, experimenters carry out Bell inequality violation
tests (c.f. Refs. \cite{Kimble, Kwiat}). Could the schemes presented here realistically
quantify entanglement in many body systems? So far we considered the entanglement
entropy in the ground state, thus implicitly assuming that the temperature is $T=0$.
Realistically, however, the temperature is finite, the state under consideration is not
in a pure state, and Eq. (\ref{measure}) should, in principle, be generalized to
account for thermal fluctuations. However, one can always imagine obtaining a lower
bound on the ground state entanglement entropy by extrapolating the appropriate
measurement entropy, as detailed above, to $T=0$. We leave the investigation of the
efficacy of such realistic procedures for future work.

We thank J. Preskill and L. B. Ioffe for discussions.



\begin{thebibliography}{99}
\bibitem{note1} Nevertheless even for observables for which
the measurement entropy is larger then the Von Neuman entropy,
information can still be gained in the measurement \cite{Balian}.
\bibitem{note2} Trivially $I\times I\in{\cal L}$. We may also
construct other operators by considering the Schmidt decomposition
$\psi=\sum \alpha_i |i_A>\otimes|i_B>$, for any couple $(\hat{O}_A
, \hat{O}_B)$ of operators diagonal in the $|i_A>,|i_B>$ bases,
and such that $<i_A|\hat{O}_A|i_A>+<i_B|\hat{O}_B|i_B>=const$, the
operator $O$ belongs to ${\cal L}$.
\bibitem{note3} For example the two spin Hamiltonian $\sigma_3\otimes
I+I\otimes \sigma_3+\sigma_1\otimes \sigma_1$, does not commute
with any observable in the appropriate $\cal L$.
\end{thebibliography}
\end{document}